\documentclass[reprint, secnumarabic,amssymb, nobibnotes, aps, prb]{revtex4-1}

\pdfoutput=0

\usepackage{color}
\usepackage{epsfig, graphics, graphicx, subfigure, wrapfig, lipsum}
\usepackage{verbatim}
\usepackage{dcolumn}
\usepackage{bm}
\usepackage{amsmath, braket}
\usepackage{array}
\usepackage{xr}

\newcolumntype{X}[1]{>{\centering\arraybackslash\hspace{0pt}}p{#1}}
\newcolumntype{M}[1]{ >{\centering\arraybackslash}m{#1}}
\newcommand{\roml}[1]{\lowercase\expandafter{\romannumeral #1\relax}}
\newcommand{\romu}[1]{\uppercase\expandafter{\romannumeral #1\relax}}

\begin{document}

\title{Accelerating Phonon Thermal Conductivity Prediction by an Order of Magnitude Through Machine Learning-Assisted Extraction of Anharmonic Force Constants}

\author{Yagyank Srivastava and Ankit Jain}
\affiliation{Mechanical Engineering Department, IIT Bombay, India}
\date{\today}%

\begin{abstract}
The calculation of material phonon thermal conductivity from density functional theory calculations requires computationally expensive evaluation of anharmonic interatomic force constants and has remained a computational bottleneck in the high-throughput discovery of materials. In this work, we present a machine learning-assisted approach for the extraction of anharmonic force constants through local learning of the potential energy surface. We demonstrate our approach on a diverse collection of 220 ternary materials for which the total computational time for anharmonic force constants evaluation is reduced by more than an order of magnitude from 480,000 cpu-hours to less than 12,000 cpu-hours while preserving the thermal conductivity prediction accuracy to within 10\%. Our approach removes a major hurdle in computational thermal conductivity evaluation and will pave the way forward for the high-throughput discovery of materials.
\end{abstract}
\maketitle

\section{Introduction}
The determination of phonon thermal conductivity ($\kappa$) is crucial in applications such as thermoelectrics, thermal barrier coating, and heat dissipation \cite{slack1995, clarke2003, minnich2009a, lindsay2018a}. The conventional search of high/low-$\kappa$ materials is via the experimental trial-and-error approach, which is slow and resource-consuming. While data-driven structure-property relation-based end-to-end machine learning (ML) approaches have been tested for $\kappa$ prediction, these approaches are shown to have insufficient accuracy, owing mainly to the non-availability of high-quality systematic $\kappa$-datasets \cite{zhu2021,liu2022,  srivastava2023}.
With recent developments in computational approaches, it is now possible to predict $\kappa$ of materials via the Boltzmann transport equation (BTE) approach with input from ab initio calculations \cite{esfarjani2008, esfarjani2011, lindsay2018, mcgaughey2019, veeravenkata2021}. This, however, requires interatomic force constants (IFCs) and phonon scattering rates, each of which, depending on the material system, requires several hundred to thousands of computational hours \cite{esfarjani2008, jain2020}.
Several ML approaches have been developed recently to accelerate the phonon scattering rates calculations \cite{guo2023, srivastava2024}. The extraction of harmonic and anharmonic IFCs has remained a computational bottleneck in the high-throughput discovery of materials.

In this work, an ML-assisted approach is presented for the extraction of IFCs from ab-initio based density functional theory (DFT) calculations. In this approach, an ML model is first trained to learn the local potential energy surface as experienced by atoms and this trained model is then used to replace DFT and predict forces needed in the IFCs extraction. With this seemingly simple approach, the computational cost of anharmonic IFCs extraction is reduced by more than an order of magnitude and the accuracy of $\kappa$ prediction is preserved to within 10\%.

The approach presented here is different than frequently used ML force field development approaches where the objective is to sample the complete potential energy surface and the development/training of forcefield itself requires computationally expensive training dataset \cite{bartok2018, mortazavi2021, yulou2022, tang2023, rodriguez2023,rodriguez2023a,lee2024}. In the approach presented here, since the objective is to learn the local potential energy surface, only a handful of DFT calculations are sufficient to train the ML model.

\section{Methodology} The contribution of phonons towards the $\kappa$ of material is obtainable using the BTE along with the Fourier's law as \cite{ziman1960, reissland1973, srivastava1990}:
\begin{equation}
 \label{eqn_theory_conduct}
 \kappa = \sum_{\lambda}  c_{\lambda} v_{\lambda, \alpha}^{2} \tau_{\lambda, \alpha},
\end{equation}
where the summation is over all phonon modes, $c_{\lambda}$ is the phonon specific heat, $v_{\lambda,\alpha}$ is the $\alpha$ component of phonon group velocity vector ${\boldsymbol{v}_{\lambda}}$, and $\tau_{\lambda, \alpha}$ is the phonon scattering time. The calculation of phonon heat capacity {\color{black} (using the Bose-Einstein statistics)}, group velocity, and scattering rate requires harmonic and anharmonic (cubic/quartic) IFCs and more details on these are presented elsewhere in \cite{mcgaughey2019, jain2020}. Depending on the material, the total number of such IFCs varies between several hundred to millions. However, since these IFCs are interrelated through underlying crystal symmetries, the number of unique symmetry-unrelated IFCs varies between several hundred to thousands for different material systems \cite{esfarjani2008}. The DFT calculations for extraction of these IFCs are typically carried on 100-400 atom computational cells to extract all needed IFCs within the given cutoff while avoiding interactions from periodic images (in case of a periodic basis set).

The straightforward approach of obtaining the symmetry-unrelated IFCs is via the finite-difference approach, where one or more atoms are displaced from their equilibrium position and the restoring forces on these perturbed configurations are obtained from DFT calculations \cite{esfarjani2008}.
With this approach, the required number of DFT calculations varies from several hundred to thousands. For instance, for Si, the number of required DFT calculations are $4$ and $52$ for harmonic and cubic IFCs with interaction cutoff of $5^{\text{th}}$ nearest neighbor, and the corresponding numbers are $18$ and $488$ for type-I $\text{Ba}_8\text{Ga}_{16}\text{Ge}_{30}$ (BGG) clathrate with a more complex crystal structure. The IFCs can also be obtained, alternatively, via the Taylor-series force-displacement dataset fitting of the over-specified system (number of force-displacement relations more than the number of unknown IFCs). This force-displacement dataset can either be sampled from molecular dynamics trajectory or could be obtained stochastically using the thermal snapshot technique \cite{west2006, hellman2013, shulumba2017}. The molecular dynamics trajectory sampling requires $\sim$10,000 DFT-MD timesteps to obtain non-correlated thermal configurations, whereas the thermal snapshot approach requires the upfront knowledge of phonon vibration spectra. Depending on the material system,  the Taylor-series-based approaches require $\sim$100-400 thermally perturbed configurations for IFCs extraction \cite{ravichandran2018, jain2020}. 

An ML-assisted approach can be explored to reduce the computational cost of DFT-based IFCs extraction (originating from force evaluations on many perturbed configurations). In the ML-assisted approach, an ML model can be first trained on actual DFT forces of a few configurations. The trained ML model can then be used to replace DFT and obtain forces on remaining configurations. Since the underlying objective is to reduce the number of DFT calculations through local learning of the potential energy surface, physics-motivated ML models, such as Gaussian Approximation potential (GAP), are expected to outperform purely data-driven ML models, like those based on neural networks, even though later are more flexible (adaptable for any generic task at the expense of large data requirement). We employed a GAP ML model with two-body, three-body, and the smooth overlap of atomic orbitals (SOAP) descriptors  \cite{bartok2010, bartok2013}. 
The cutoffs employed for these descriptors are set at $5.0$, $4.0$, and $5.0$ $\text{\AA}$ and the number of angular/radial symmetry functions in SOAP descriptor is set at $6$/$12$ for each atomic species. The training is performed on DFT forces using Quantum Mechanics and Interatomic Potentials (QUIP) code with regularization of  $0.2$ \cite{bartok2010}. 
{\color{black} The complete computational workflow with an ML-assisted approach is reported in the S1 in the SM. The details of the force-displacement dataset fitting for interatomic force constants extraction are the same as those reported in Ref.~\cite{srivastava2023, srivastava2024}. Since the focus here is on the acceleration of the anharmonic interatomic force constants extraction, only the relevant methodology is presented. The details on the iterative solution of the BTE, the four-phonon scatterings, the contribution of coherent wave-like transport channels, etc, are reported elsewhere in Refs.~\cite{mukhopadhyay2018, simoncelli2019, leyla2019, giovanni2022, jain2022, alfredo2023, jain2024}.}

\section{Results} 
\begin{figure}
\begin{center}
\epsfbox{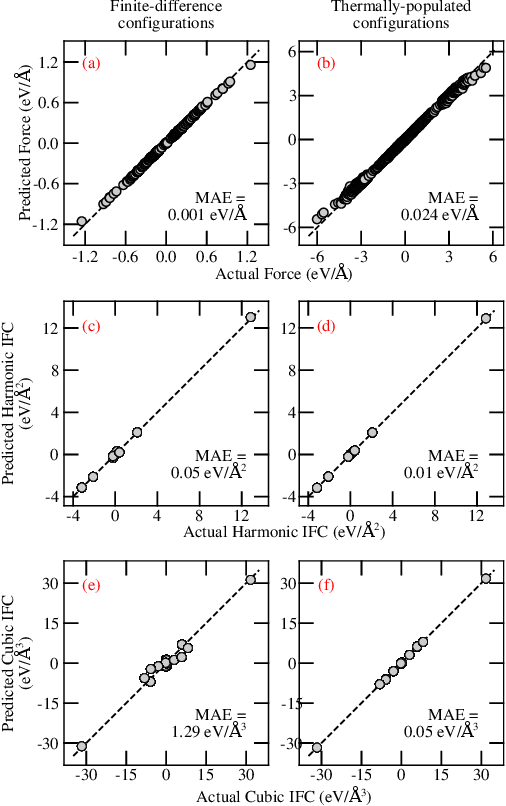}
\end{center}
\caption{The performance of trained ML models for (a), (b) force prediction, (c), (d) harmonic IFC prediction, and (e), (f) cubic IFC prediction of silicon. The model in (a), (c), (e) is trained on finite-difference perturbation-based configurations and the model in (b), (d), (f) is trained on thermally populated configurations. The reported forces in (a), (b) are on test datasets consisting of 42 and 190 configurations, respectively. While both models are able to predict forces and harmonic IFCs with sufficient accuracy, the cubic IFCs obtained from the later model, which is trained on thermally populated configurations, result in an order of magnitude less MAE compared to that from the former model (trained on finite-difference based perturbations).
}
\label{fig_Si}
\end{figure}

We start by exploring the possibility of ML-assisted IFC extraction for Si at 300 K based on the finite-difference approach. 
We mixed the harmonic and anharmonic IFC configurations and trained the ML model on 25\% (14) of a total of $56$ configurations. When tested on the remaining 75\% configurations, we find that the trained ML model results in an excellent force prediction performance with mean absolute error (MAE) of only 1 meV/$\text{\AA}$ [Fig.~\ref{fig_Si}(a)]. When we used these predicted forces to extract IFCs, however, we found that while harmonic IFCs were in agreement with actual harmonic IFCs (MAE of $0.05$ $\text{eV}/\text{\AA}^2$), the cubic IFCs had larger variations with MAE of $1.29$ $\text{eV}/\text{\AA}^3$ [Figs.~\ref{fig_Si}(c), \ref{fig_Si}(e)]. 
We believe that owing to a small magnitude of forces in finite-difference-based approach (due to small perturbations of atoms by $0.005$-$0.05$ $\text{\AA}$), the ML model is able to get the bulk part right (MAE of only 1 meV/$\text{\AA}$ on predicted forces and $0.05$ $\text{eV}/\text{\AA}^2$ on predicted harmonic IFCs), but failed in distinguishing numerical noise from anharmonic force contribution.

Consequently, we computed phonon vibration spectra using the actual DFT forces-based harmonic IFCs and employed it with the thermal snapshot technique to obtain 200 thermally populated configurations (all atoms are displaced in the computational cell corresponding to the thermal population of phonon normal modes) \cite{west2006}. We employed 10 such thermal configurations for ML training and predicted forces on the remaining configurations for model performance evaluation. The forces obtained on 190 test configurations using this trained ML model are plotted against the actual DFT forces in Fig.~\ref{fig_Si}(b). 

At first, the MAE obtained on predicted forces seems high from the thermal snapshot trained model compared to that obtained for the finite-difference trained model. In the thermal snapshot dataset, however, the force magnitude is also large owing to larger thermal displacements/perturbations of atoms. As such, MAE on predicted forces is not a fair comparison matrix for ML models in this scenario. IFCs, on the other hand, are independent of atomic perturbations and are expected to have a similar range from different extraction approaches. The IFCs obtained using the thermal snapshot-trained model are reported in  Figs.~\ref{fig_Si}(d) and \ref{fig_Si}(f). We find that the prediction performance of the thermal snapshot-trained model is superior compared to the finite-difference-trained model for both harmonic and cubic IFCs. In particular, the MAE on extracted cubic IFCs is an order of magnitude less with the thermal snapshot trained model and is only $0.15$ $\text{eV}/\text{\AA}^3$. 

\begin{figure}
\begin{center}
\epsfbox{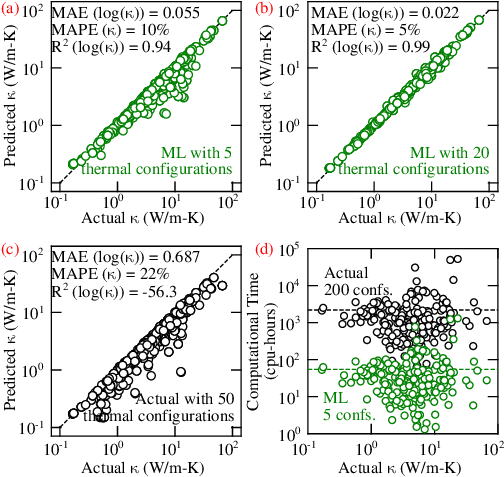}
\end{center}
\caption{The machine learning model performance for thermal conductivity prediction of 220 diverse ternary materials.  The $\kappa$ obtained with (a), (b) ML predicted forces with 5 and 20 thermal configurations in the training dataset and (c) without ML assistance with 50 thermal configurations in the force-displacement dataset fitting. (d) The DFT computational cost associated with anharmonic IFCs extraction with 5 configurations in the ML training dataset. The dashed horizontal lines in (d) denote the average anharmonic IFC extraction computational time of considered materials which is reduced by more than an order of magnitude with ML assistance.}
\label{fig_ML}
\end{figure}

Motivated by this, we next test if this impressive performance of the thermal snapshot-trained ML model is limited to Si or if it is transferable to other materials. For this, we considered a diverse set of 220 ternary materials for which the $\kappa$ are obtained using systematic high-throughput DFT calculations and span around three orders of magnitude as detailed in Ref.~\cite{srivastava2023}. For each considered material, we trained a separate ML model and used the trained model to predict forces on 200 thermal configurations. We test the performance of the trained ML model to predict $\kappa$ by employing 5 and 20 configurations in the training dataset for ML models.
The $\kappa$ obtained using these trained ML model-based IFCs are reported in Figs.~\ref{fig_ML}(a) and \ref{fig_ML}(b). 

As can be seen from Fig.~\ref{fig_ML}, we find that the ML-assisted model is able to predict $\kappa$ of diverse materials surprisingly well with mean absolute percentage error (MAPE) of $10\%$ with only 5 configurations in the ML training dataset. Further, with the inclusion of 10 and 20 structures in the training dataset, 
the MAPE reduces to 7\% and 5\%, respectively. In comparison,
when we extract IFCs without employing the ML model [Fig.~\ref{fig_ML}(c], the MAPE stays high at $22\%$ even with 50 configurations in the Taylor-series force-displacement dataset fitting; thus, clearly highlighting the assistive role of ML model in learning the local potential energy surface for the IFCs extraction. 

In Fig.~\ref{fig_ML}(d), we report the computational time involved in the extraction of anharmonic IFCs with and without an ML-assisted approach for considered 220 materials with 5 configurations in the training dataset. The total DFT computational cost for anharmonic IFCs evaluation for considered 220 materials is 482,000 cpu-hours, and it is reduced to less than 12,000 cpu-hours with an ML-assisted approach; thus, reflecting an order of magnitude reduction in the anharmonic IFCs evaluation computational cost with ML-assisted approach.

\begin{figure}
\begin{center}
\epsfbox{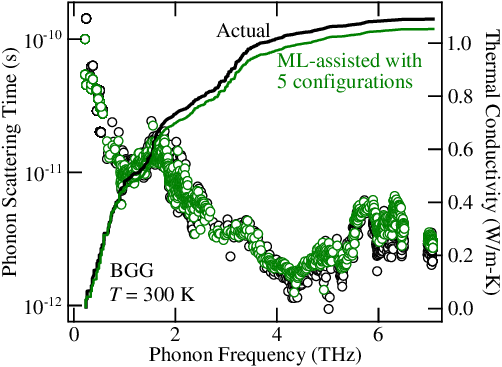}
\end{center}
\caption{The performance of ML-assisted approach for thermal transport properties prediction of type-I BGG clathrate. The total end-to-end computational cost for $\kappa$ prediction reduces by more than a factor of nine with an ML-assisted approach while preserving the accuracy of predicted $\kappa$ to within 4\%. }
\label{fig_BGG}
\end{figure}

We note that the ML-assisted approach presented here for IFCs extraction is complimentary to the ML-accelerated approach for phonon scattering rates calculations \cite{srivastava2024}. These two approaches can be used in conjunction to reduce the total computational cost of $\kappa$ prediction. To showcase the cumulative effect of these ML-based approaches, we consider thermal transport in BGG type-I clathrate with a complex crystal structure \cite{godse2022}. Without any ML, the total computational time to obtain $\kappa$ of BGG is around 47,000 cpu-hours Fig.~\ref{fig_BGG} of which $8$\% and 90\% are spent respectively on DFT calculations for harmonic and anharmonic IFCs (with 200 thermal configurations) and $1$\% on phonon scattering rates calculations.
With an ML-assisted approach for anharmonic IFCs extraction, 
the computational time for DFT calculations reduces from more than 42,000 cpu-hours to less than 1100 cpu-hours and with ML-accelerated phonon scattering rates calculations,  the computational time for phonon-scattering rates calculations reduces from 550 cpu-hours to 15 cpu-hours. Consequently, the total end-to-end computational time reduces from 47,000 to 5000 cpu-hours with ML-assisted approaches for BGG. Note that, for BGG, around 100 cpu-hours are spent on miscellaneous tasks such as dynamical matrix diagonalization, force-displacement data-fitting, etc, which remains unchanged with the ML approach. Further, the computational time for ML training is only $0.5$ cpu-hours for BGG.

The $\kappa$ obtained using the ML-assisted approach for BGG is $1.05$ W/m-K at 300 K, which is within 4\% of that obtained using the actual calculations. Further, the obtained mode-dependent phonon properties from the ML approach (Fig.~\ref{fig_BGG}) show minimal deviation from the actual values; thus, further highlighting the role of ML in accelerating $\kappa$ calculations while preserving the prediction accuracy.

\begin{figure}
\begin{center}
\epsfbox{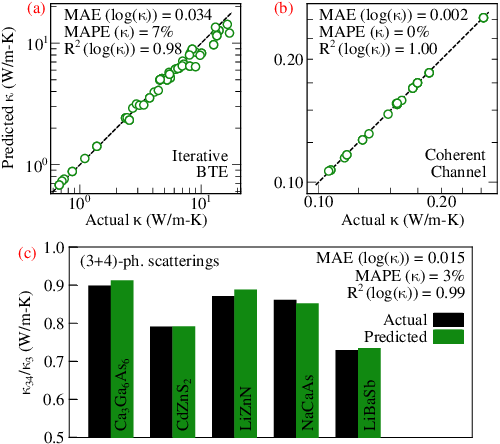}
\end{center}
\caption{{\color{black} The performance of ML-assisted approach in describing the higher-order thermal transport physics: (a) $\kappa$ obtained using the iterative solution of the BTE, (b) contribution of the wave-like coherent transport channel to $\kappa$, and (c) $\kappa$ obtained by including the four-phonon scattering. The ML model is trained using 20 thermal configurations for each of the considered materials.}}
\label{fig_higherorder}
\end{figure}

\subsection{Higher-order thermal transport physics}
So far, we have tested the applicability of our ML-assisted IFCs extraction approach only for three-phonon scattering driven $\kappa$ calculation under the relaxation time approximation of the BTE with the particle-like single-transport channel. Recent studies demonstrated the need of higher-order thermal transport physics involving iterative/full solution of the BTE, four-phonon scattering, and multi-channel thermal transport for various material systems {\color{black}\cite{mukhopadhyay2018, simoncelli2019, leyla2019, giovanni2022, jain2022, alfredo2023, jain2024}}.

To test the applicability of the proposed ML-assisted approach in describing these higher-order thermal transport physics, we randomly chose 10\% of the considered ternary materials and carried out $\kappa$ calculations using the full/iterative solution of the BTE and by accounting for the contribution of the wave-like coherent transport channel. For four-phonon scattering, since the associated computational cost of phonon scattering rate calculation is itself high, we randomly chose five materials from the dataset.

We find that with the iterative solution of the BTE, the ML-assisted IFCs-based $\kappa$ have MAPE of only 7\% [Fig.~\ref{fig_higherorder}(a)] and for the coherent channel thermal transport, the ML-assisted IFCs-based $\kappa$ are in perfect agreement with the actual values [Fig.~\ref{fig_higherorder}(b)]. Further, for four-phonon scattering, the $\kappa$ obtained using the ML-assisted approach have MAPE of only 3\%; thus, clearly demonstrating the applicability of the proposed approach for the description of higher-order thermal transport physics.

\subsection{Comparison with other ML approaches}
The SOAP-GAP ML model employed here has also been used in many literature studies to learn the potential energy surface. However, as opposed to the learning of the local potential energy surface, the literature studies have focused on global force field training through complete training of the potential energy surface, requiring a much larger training dataset with several hundred to thousands of atomic configurations \cite{babaei2019, korotaev2019}. Further, many approaches are reported in the literature for force field training, including hiPhive (employing advanced optimization techniques to find sparse solutions \cite{Eriksson2019-oh}) and MLIP (employing moment tensor potentials \textcolor{black}{MTPs} \cite{Novikov2021}). {\color{black}The major differences between these approaches are highlighted in Section S2 of SM.}

\begin{figure}
\begin{center}
\epsfbox{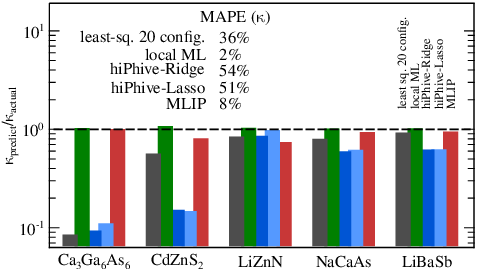}
\end{center}
\caption{\color{black} The performance comparison of localized ML-assisted approaches with other ML/regularization approaches. All models are trained using the same 20 atomic configurations. The ML-assisted approaches outperform other considered approaches.}
\label{fig_diffmodels}
\end{figure}

We compare the performance of hiPhive and MLIP with a  SOAP-GAP model in Fig.~\ref{fig_diffmodels}. For this comparison, we consider the same five materials as those employed in Fig.~\ref{fig_higherorder}(c). With hiPhive, we employ $6.5$ $\text{\AA }$ interaction cutoff and test two different penalty functions: Ridge penalty (based on L2 norm) and Lasso penalty (based on L1 norm) while keeping other parameters at their default settings. For MLIP, we use moment tensor potential of level 8 with a radial basis of size 8 {\color{black}with 1000 iterations}. We note that hiPhive with the least-square penalty is the same as an actual model (no ML/regularization), requiring a large dataset ($\sim$200 thermal configurations) [see Fig.~\ref{fig_ML}(c)].

We find that for considered materials, when 20 atomic configurations are used, and force constants are extracted without any ML assistance using the least-square fitting, the system is weakly over-determined (the number of unknown force constants is similar to a number of force-displacement relations), and the obtained MAPE is high at 36\%. When these same 20 configurations are used for the extraction of force constants using the L1- or L2-norm regularization in hiPhive, the error in $\kappa$ prediction increases to more than 50\%, which is expected due to the inclusion of an additional fitting penalty in weakly over-determined systems. To make systems over-determined, when forces are obtained on 200 atomic configurations using the MLIP model trained on 20 atomic configurations, {\color{black}the MAPE on obtained $\kappa$ is only 8\%}. Similarly, when forces on the same 200 atomic configurations are obtained using the SOAP GAP model (trained on the same 20 atomic configurations), the obtained MAPE is only 2\%. 
These obtained MAPE with the locally-trained ML models are much better than that without any ML assistance, thus clearly reflecting the advantage of the proposed approach in force constants extraction of materials.

\section{Conclusions}
In summary, we explored the possibility of using an ML approach for accelerated evaluation of anharmonic force constants which are needed for the computation of phonon thermal conductivity in materials. For the considered SOAP-GAP-based locally trained ML model, we find that while predicted interatomic forces and extracted harmonic force constants are accurate when trained on finite-difference-based perturbations, the anharmonic force constants are less accurate. On the other hand, when the same ML model is trained on thermally perturbed structures, the obtained anharmonic force constants are also accurate. 
The accuracy of thermal conductivity obtained from these ML-driven anharmonic force constants is preserved to within 10\% while the computational cost is reduced by an order of magnitude. {\color{black}While the present work focussed only on anharmonic force constants, efforts are currently underway to extend it to harmonic force constants.} Our proposed approach reduces the computational cost of DFT-driven thermal conductivity prediction and presents a way forward for the high-throughput discovery of materials.

\section{Acknowledgement}
The authors acknowledge the financial support from the Core Research Grant, Science \& Engineering Research Board, India (Grant Number: CRG/2021/000010) and Nano
Mission, Government of India (Grant No.: DST/NM/NS/2020/340).  The calculations are carried out on the SpaceTime-II supercomputing facility of IIT Bombay. 

\section{Data Availability}
{\color{black} The following data is available in the Supplementary Material: (a) 200 thermally perturbed structures of each considered compound, (b) DFT forces on 200 thermally perturbed structures of each considered compound, and (c) scripts to perform ML using SOAP GAP, MTP MLIP, and hiPhive models. }


\begin{thebibliography}{10}

\bibitem{slack1995}
G.~A. Slack.
\newblock New materials and performance limits for thermoelectric cooling.
\newblock In D.~M. Rowe, editor, {\em CRC Handbook of Thermoelectrics}, pages 407--440. CRC, Boca Raton, 1995.

\bibitem{clarke2003}
David~R Clarke.
\newblock Materials selection guidelines for low thermal conductivity thermal barrier coatings.
\newblock {\em Surface and Coatings Technology}, 163:67--74, 2003.

\bibitem{minnich2009a}
A.~J. Minnich, M.~S. Dresselhaus, F.~Ren, and G.~Chen.
\newblock Bulk nanostructured thermoelectric materials: current research and future prospects.
\newblock {\em Energy and Environmental Sciences}, 2:466--479, 2009.

\bibitem{lindsay2018a}
Lucas Lindsay and Carlos~A. Polanco.
\newblock {\em {Thermal Transport by First-Principles Anharmonic Lattice Dynamics}}.
\newblock 2018.

\bibitem{zhu2021}
Taishan Zhu, Ran He, Sheng Gong, Tian Xie, Prashun Gorai, Kornelius Nielsch, and Jeffrey~C Grossman.
\newblock Charting lattice thermal conductivity for inorganic crystals and discovering rare earth chalcogenides for thermoelectrics.
\newblock {\em Energy \& Environmental Science}, 14(6):3559--3566, 2021.

\bibitem{liu2022}
Zeyu Liu, Meng Jiang, and Tengfei Luo.
\newblock Leveraging low-fidelity data to improve machine learning of sparse high-fidelity thermal conductivity data via transfer learning.
\newblock {\em Materials Today Physics}, 28:100868, 2022.

\bibitem{srivastava2023}
Yagyank Srivastava and Ankit Jain.
\newblock End-to-end material thermal conductivity prediction through machine learning.
\newblock {\em Journal of Applied Physics}, 134(22), December 2023.

\bibitem{esfarjani2008}
K.~Esfarjani and H.~T. Stokes.
\newblock Method to extract anharmonic force constants from first principles calculations.
\newblock {\em Physical Review B}, 77:144112, 2008.

\bibitem{esfarjani2011}
K.~Esfarjani, G.~Chen, and H.~T. Stokes.
\newblock Heat transport in silicon from first-principles calculations.
\newblock {\em Physical Review B}, 84:085204, 2011.

\bibitem{lindsay2018}
Lucas Lindsay, Chengyun Hua, XL~Ruan, and Sangyeop Lee.
\newblock Survey of ab initio phonon thermal transport.
\newblock {\em Materials Today Physics}, 7:106--120, 2018.

\bibitem{mcgaughey2019}
Alan~J.H. McGaughey, Ankit Jain, Hyun~Young Kim, and Bo~Fu.
\newblock {Phonon properties and thermal conductivity from first principles, lattice dynamics, and the Boltzmann transport equation}.
\newblock {\em Journal of Applied Physics}, 125(1):11101, jan 2019.

\bibitem{veeravenkata2021}
Harish~P Veeravenkata and Ankit Jain.
\newblock Density functional theory driven phononic thermal conductivity prediction of biphenylene: A comparison with graphene.
\newblock {\em Carbon}, 183:893--898, 2021.

\bibitem{jain2020}
Ankit Jain.
\newblock Multichannel thermal transport in crystalline {Tl$_3$VSe$_4$}.
\newblock {\em Physical Review B}, 102(20), November 2020.

\bibitem{guo2023}
Ziqi Guo, Prabudhya Roy~Chowdhury, Zherui Han, Yixuan Sun, Dudong Feng, Guang Lin, and Xiulin Ruan.
\newblock Fast and accurate machine learning prediction of phonon scattering rates and lattice thermal conductivity.
\newblock {\em npj Computational Materials}, 9(1):95, 2023.

\bibitem{srivastava2024}
Yagyank Srivastava and Ankit Jain.
\newblock Accelerating thermal conductivity prediction through machine-learning: Two orders of magnitude reduction in phonon-phonon scattering rates calculation.
\newblock {\em Materials Today Physics}, 41:101345, February 2024.

\bibitem{bartok2018}
Albert~P. Bart\'ok, James Kermode, Noam Bernstein, and G\'abor Cs\'anyi.
\newblock Machine learning a general-purpose interatomic potential for silicon.
\newblock {\em Phys. Rev. X}, 8:041048, Dec 2018.

\bibitem{mortazavi2021}
Bohayra Mortazavi, Evgeny~V Podryabinkin, Ivan~S Novikov, Timon Rabczuk, Xiaoying Zhuang, and Alexander~V Shapeev.
\newblock Accelerating first-principles estimation of thermal conductivity by machine-learning interatomic potentials: A mtp/shengbte solution.
\newblock {\em Computer Physics Communications}, 258:107583, 2021.

\bibitem{yulou2022}
Yulou Ouyang, Cuiqian Yu, Jia He, Pengfei Jiang, Weijun Ren, and Jie Chen.
\newblock Accurate description of high-order phonon anharmonicity and lattice thermal conductivity from molecular dynamics simulations with machine learning potential.
\newblock {\em Phys. Rev. B}, 105:115202, Mar 2022.

\bibitem{tang2023}
Jialin Tang, Guotai Li, Qi~Wang, Jiongzhi Zheng, Lin Cheng, and Ruiqiang Guo.
\newblock Competition between phonon-vacancy and four-phonon scattering in cubic boron arsenide by machine learning interatomic potential.
\newblock {\em Phys. Rev. Mater.}, 7:044601, Apr 2023.

\bibitem{rodriguez2023}
Alejandro Rodriguez, Changpeng Lin, Chen Shen, Kunpeng Yuan, Mohammed Al-Fahdi, Xiaoliang Zhang, Hongbin Zhang, and Ming Hu.
\newblock Unlocking phonon properties of a large and diverse set of cubic crystals by indirect bottom-up machine learning approach.
\newblock {\em Communications Materials}, 4(1):61, 2023.

\bibitem{rodriguez2023a}
Alejandro Rodriguez, Changpeng Lin, Hongao Yang, Mohammed Al-Fahdi, Chen Shen, Kamal Choudhary, Yong Zhao, Jianjun Hu, Bingyang Cao, Hongbin Zhang, et~al.
\newblock Million-scale data integrated deep neural network for phonon properties of heuslers spanning the periodic table.
\newblock {\em npj Computational Materials}, 9(1):20, 2023.

\bibitem{lee2024}
Huiju Lee and Yi~Xia.
\newblock Machine learning a universal harmonic interatomic potential for predicting phonons in crystalline solids.
\newblock {\em Applied Physics Letters}, 124(10), 2024.

\bibitem{ziman1960}
J.~M. Ziman.
\newblock {\em Electrons and Phonons}.
\newblock Oxford University Press, Clarendon, Oxford, 1960.

\bibitem{reissland1973}
J.~A. Reissland.
\newblock {\em The Physics of Phonons}.
\newblock John Wiley and Sons Ltd, 1973.

\bibitem{srivastava1990}
G.~P. Srivastava.
\newblock {\em The Physics of Phonons}.
\newblock Adam Hilger, Bristol, 1990.

\bibitem{west2006}
D~West and SK~Estreicher.
\newblock First-principles calculations of vibrational lifetimes and decay channels: Hydrogen-related modes in si.
\newblock {\em Physical review letters}, 96(11):115504, 2006.

\bibitem{hellman2013}
Olle Hellman, Peter Steneteg, Igor~A Abrikosov, and Sergei~I Simak.
\newblock Temperature dependent effective potential method for accurate free energy calculations of solids.
\newblock {\em Physical Review B}, 87(10):104111, 2013.

\bibitem{shulumba2017}
Nina Shulumba, Olle Hellman, and Austin~J Minnich.
\newblock Intrinsic localized mode and low thermal conductivity of pbse.
\newblock {\em Physical Review B}, 95(1):014302, 2017.

\bibitem{ravichandran2018}
Navaneetha~K Ravichandran and David Broido.
\newblock Unified first-principles theory of thermal properties of insulators.
\newblock {\em Physical Review B}, 98(8):085205, 2018.

\bibitem{bartok2010}
Albert~P. Bart\'ok, Mike~C. Payne, Risi Kondor, and G\'abor Cs\'anyi.
\newblock Gaussian approximation potentials: The accuracy of quantum mechanics, without the electrons.
\newblock {\em Phys. Rev. Lett.}, 104:136403, Apr 2010.

\bibitem{bartok2013}
Albert~P. Bart\'ok, Risi Kondor, and G\'abor Cs\'anyi.
\newblock On representing chemical environments.
\newblock {\em Phys. Rev. B}, 87:184115, May 2013.

\bibitem{mukhopadhyay2018}
Saikat Mukhopadhyay, David~S Parker, Brian~C Sales, Alexander~A Puretzky, Michael~A McGuire, and Lucas Lindsay.
\newblock Two-channel model for ultralow thermal conductivity of crystalline $\text{Tl}_3\text{VSe}_4$.
\newblock {\em Science}, 360(6396):1455--1458, 2018.

\bibitem{simoncelli2019}
Michele Simoncelli, Nicola Marzari, and Francesco Mauri.
\newblock Unified theory of thermal transport in crystals and glasses.
\newblock {\em Nature Physics}, 15(8):809--813, 2019.

\bibitem{leyla2019}
Leyla Isaeva, Giuseppe Barbalinardo, Davide Donadio, and Stefano Baroni.
\newblock Modeling heat transport in crystals and glasses from a unified lattice-dynamical approach.
\newblock {\em Nat. Commun.}, 10(1):3853, August 2019.

\bibitem{giovanni2022}
Giovanni Caldarelli, Michele Simoncelli, Nicola Marzari, Francesco Mauri, and Lara Benfatto.
\newblock Many-body green's function approach to lattice thermal transport.
\newblock {\em Phys. Rev. B}, 106:024312, Jul 2022.

\bibitem{jain2022}
Ankit Jain.
\newblock Single-channel or multichannel thermal transport: Effect of higher-order anharmonic corrections on the predicted phonon thermal transport properties of semiconductors.
\newblock {\em Phys. Rev. B}, 106:045207, Jul 2022.

\bibitem{alfredo2023}
Alfredo Fiorentino and Stefano Baroni.
\newblock From green-kubo to the full boltzmann kinetic approach to heat transport in crystals and glasses.
\newblock {\em Phys. Rev. B}, 107:054311, Feb 2023.

\bibitem{jain2024}
Ankit Jain.
\newblock Impact of four-phonon scattering on thermal transport in carbon nanotubes.
\newblock {\em Physical Review B}, 109(15):155413, 2024.

\bibitem{godse2022}
Shravan Godse, Yagyank Srivastava, and Ankit Jain.
\newblock Anharmonic lattice dynamics and thermal transport in type-i inorganic clathrates.
\newblock {\em Journal of Physics: Condensed Matter}, 34(14):145701, February 2022.

\bibitem{babaei2019}
Hasan Babaei, Ruiqiang Guo, Amirreza Hashemi, and Sangyeop Lee.
\newblock Machine-learning-based interatomic potential for phonon transport in perfect crystalline si and crystalline si with vacancies.
\newblock {\em Phys. Rev. Mater.}, 3:074603, Jul 2019.

\bibitem{korotaev2019}
Pavel Korotaev, Ivan Novoselov, Aleksey Yanilkin, and Alexander Shapeev.
\newblock Accessing thermal conductivity of complex compounds by machine learning interatomic potentials.
\newblock {\em Phys. Rev. B}, 100:144308, Oct 2019.

\bibitem{Eriksson2019-oh}
Fredrik Eriksson, Erik Fransson, and Paul Erhart.
\newblock The hiphive package for the extraction of high‐order force constants by machine learning.
\newblock {\em Adv. Theory Simul.}, 2(5):1800184, May 2019.

\bibitem{Novikov2021}
Ivan~S Novikov, Konstantin Gubaev, Evgeny~V Podryabinkin, and Alexander~V Shapeev.
\newblock The mlip package: moment tensor potentials with mpi and active learning.
\newblock {\em Machine Learning: Science and Technology}, 2(2):025002, January 2021.

\end{thebibliography}

\end{document}